\documentstyle[epsfig]{aipproc}

\newcommand{\alt}{\raisebox{-0.13cm}{~\shortstack{$<$ \\[-0.07cm] $\sim$}}~}
\newcommand{\agt}{\raisebox{-0.13cm}{~\shortstack{$>$ \\[-0.07cm] $\sim$}}~}

\begin{document}

\begin{flushright}
{\bf MADPH-98-1043}, {\bf hep-ph/9802343}
\end{flushright}

\title{The Search for Higgs Bosons \\ 
of Minimal Supersymmetry at the LHC}
\thanks{
Presented at the Workshop on Physics at the First Muon Collider 
and at the Front End of a Muon Collider, 
Fermi National Accelerator Laboratory, Batavia, Illinois, 
November 6-9, 1997.}

\author{Chung Kao}
\address{Department of Physics, University of Wisconsin \\
Madison, Wisconsin 53706}


\maketitle

\begin{abstract}

The prospects for discovering neutral Higgs bosons 
in the minimal supersymmetric model (MSSM) 
and in the minimal supergravity model (MSUGRA) 
at the LHC are investigated. 
Two special discovery channels are discussed: 
(i) the photon pair decay of the MSSM CP-odd Higgs boson, and 
(ii) the muon pair decays of neutral Higgs bosons and in the MSUGRA.

\end{abstract}

\section*{Introduction}

In the minimal supersymmetric model (MSSM) \cite{MSSM}, 
there are two Higgs doublets $\phi_1$ and $\phi_2$ 
coupling to fermions with $t_3 = -1/2$ and $t_3 = +1/2$ 
respectively \cite{Guide}. 
After spontaneous symmetry breaking, there remain five physical Higgs bosons:
a pair of singly charged $H^{\pm}$,
two neutral CP-even $H^0$ (heavier) and $h^0$ (lighter),
and a neutral CP-odd $A^0$.
The Higgs potential is constrained by supersymmetry 
such that all tree-level Higgs boson masses and couplings 
are determined by two independent parameters,  
commonly chosen to be mass of the CP-odd pseudoscalar ($m_A$) 
and ratio of the vacuum expectation values (VEVs) of Higgs fields 
($\tan\beta \equiv v_2/v_1$). 

Extensive studies have been made for the detection of MSSM Higgs bosons 
at the CERN LHC 
\cite{HGG,Neutral,KZ,Z2Z2,CMS,ATLAS,ATLAS2,Review}.
Most studies have focused on the SM decay modes 
$\phi \to \gamma\gamma$ ($\phi = H^0, h^0$ or $A^0$) 
and $\phi \to ZZ$ or $ZZ^*\to 4l$ ($\phi = H^0$ or $h^0$). 
For $\tan\beta$ close to one, 
the detection modes $A^0 \to Zh^0 \to l^+l^- b\bar{b}$ 
or $l^+l^- \tau\bar{\tau}$ \cite{AZh} 
and $H^0 \to h^0 h^0 \to \gamma\gamma b\bar{b}$ \cite{ATLAS2} 
may provide channels to simultaneously discover 
two Higgs bosons of the MSSM. 
For large $\tan\beta$, 
the $\tau\bar{\tau}$ decay mode \cite{KZ,CMS,ATLAS,ATLAS2} 
is a promising discovery channel for the $A^0$ and the $H^0$; 
neutral Higgs bosons might be observable via 
their $b\bar{b}$ decays \cite{DGV,hbb}.
In some regions of parameter space, the rates for Higgs boson
decays to SUSY particles are dominant.
While these decays reduce rates for the standard modes, 
they might also open up new promising modes for Higgs detection \cite{Z2Z2}.
Recently, the muon pair decay mode was proposed \cite{Nikita,CMS,ATLAS2} 
to be a promising discovery channel for neutral Higgs bosons. 
For large $\tan\beta$, the muon pair discovery mode might 
be the only channel that allows precise reconstruction 
of Higgs masses at the LHC. 

In this article, the prospects for discovering neutral Higgs bosons 
in the MSSM and in the minimal supergravity model (MSUGRA) 
at the LHC are investigated. 
Two special discovery channels are discussed: 
(i) the search for the MSSM CP-odd Higgs boson 
via its photon pair decay \cite{AGG}, and 
(ii) the dectection of neutral Higgs bosons via their muon pair decays 
in the MSSM \cite{Nikita} and in the MSUGRA \cite{Hmm}.

\section*{The Photon Pair Discovery Channel}

In this section, we present a realistic study 
for the observability of the MSSM CP-odd Higgs boson ($A^0$) 
via its photon pair decay mode\footnote{
This important channel was not included 
in the CMS and the ATLAS technical proposals \cite{CMS,ATLAS}.} 
($A^0 \to \gamma\gamma$) with the CMS detector performance \cite{AGG}.
The cross section for the process of $pp \to A^0 \to \gamma\gamma+X$ 
is evaluated from the cross section $\sigma(pp \to A^0 +X)$ 
multiplied with the branching fractions of $A^0 \to \gamma\gamma$.
We take $m_{\tilde{q}} = m_{\tilde{g}} = \mu =$ 1000 GeV.

\begin{figure}[h] 
\centerline{\epsfig{file=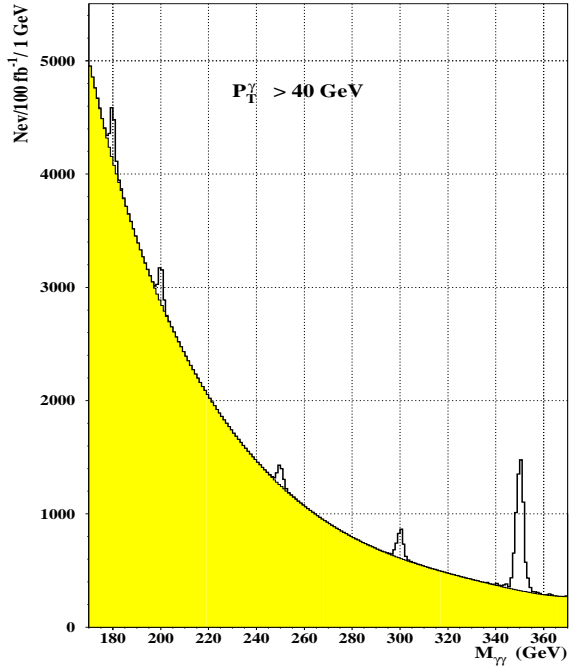,height=3.5in,width=3.0in}}
\vspace{10pt}
\caption{
Number of events versus $M_{\gamma\gamma}$, 
generated from a simulation with CMS performance, 
for the signal and the background at $\sqrt{s} = 14$ TeV 
with $L = 100$ fb$^{-1}$ and $\tan\beta = 1$.
}
\label{figure1}
\end{figure}

The irreducible backgrounds considered are, 
(i) $q \bar{q} \to\gamma \gamma $ and (ii) $ g g \to\gamma \gamma $ (Box).
In addition, we consider reducible backgrounds with at least one $\gamma$ 
in the final state, (i) $q \bar{q} \to g \gamma$, (ii)  $q g \to q \gamma$,  
and (iii) $g g \to g \gamma$ (Box). 
In Figure 1, we present number of events for the signal and the background 
at the LHC versus $M_{\gamma\gamma}$.

We use PYTHIA 5.7 and  JETSET 7.4  generators \cite{PYTHIA} 
to simulate events at the particle level.
The PYTHIA/JETSET outputs are processed
with the CMSJET program \cite{CMSJET}. 
The resolution effects are taken into account
by using the parameterizations obtained
from the detailed GEANT \cite{GEANT} simulations. 
The ECAL resolution is assumed to be 
$\sigma(E)/E = 5\% /\sqrt{E} + 0.5\%$ (CMS high luminosity regime). 
We require that every photon should have a transverse momentum ($p_T$) 
larger than 40 GeV and $|\eta| < 2.5$, 
and both photons must be isolated, {\it i.e.}, 
(i) there is no charged particle with $p_T > $ 2 GeV 
 in the cone $R = 0.3$; and 
(ii) the total transverse energy $\sum E_T^{cell}$ 
is taken to be less than 5 GeV in the cone ring 0.1 $< R < 0.3$.
To be conservative, we assume no rejection power against $\pi^0$'s 
with high $p_T$, {\it i.e.}, 
all $\pi^0$'s surviving the cuts ($p_T$, isolation, etc.) 
are considered as $\gamma$'s.
\footnote{The background from the $\pi^0$ is overestimated, especially  
in the low mass $M_{\gamma \gamma}$ region.} 

For each $m_A$ and $\tan \beta$, 
the values of mass window around the peak (within the range 2-6 GeV) 
and $p_T$ cut (50-100 GeV) 
were chosen to provide the best value of  $N_S = S/\sqrt{B}$. 
For example, the best values of the 
mass window and $p_T$ cut for  $m_A =$ 200 GeV are 2 GeV and 60 GeV 
respectively, whereas these values equal to 4 GeV and 100 GeV 
for  $m_A =$ 350 GeV. 
Figures 2 shows the discovery contour for $pp \to A^0 \to \gamma\gamma$ 
at $\sqrt{s} = 14$ TeV, in the ($m_A$,$\tan\beta$) plane,
with an integrated luminosity ($L$) of 100 fb$^{-1}$ and 300 fb$^{-1}$.

\begin{figure}[h] 
\centerline{\epsfig{file=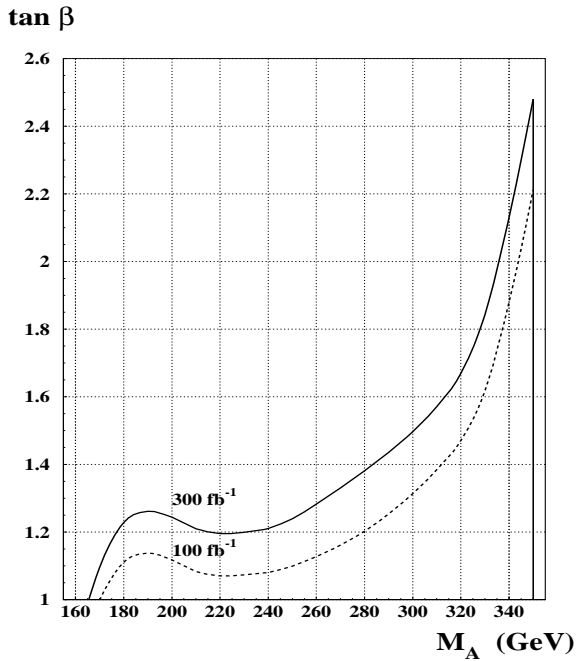,height=3.5in,width=3.0in}}
\vspace{10pt}
\caption{
The $5\sigma$ contour in the ($m_A$,$\tan\beta$) plane, 
generated from a simulation with CMS performance, 
for $pp \to A^0 \to \gamma\gamma +X$ at the LHC 
with $L = 100$ fb$^{-1}$ and 300 fb$^{-1}$.
}
\label{figure2}
\end{figure}

\section*{The Muon Pair Discovery Channel}

The cross section of $pp \to \phi \to \mu\bar{\mu} +X$ 
($\phi = A^0, H^0$, or $h^0$) 
is evaluated from the Higgs boson cross section $\sigma(pp \to \phi +X)$ 
multiplied with the branching fraction of the Higgs decay into muon pairs 
$B(\phi \to \mu\bar{\mu})$. 
The Higgs masses and couplings are evaluated with one loop corrections 
from the top and the bottom Yukawa interactions 
in the one-loop effective potential \cite{Higgs}.  


In the MSSM, gluon fusion ($gg \to \phi$) is the major source 
of neutral Higgs bosons for $\tan\beta \alt$ 4.
If $\tan\beta$ is larger than about 10,
neutral Higgs bosons are dominantly produced
from $b$-quark fusion ($b\bar{b} \to \phi$) \cite{Duane}
because the $\phi b\bar{b}$ couplings are enhanced by $1/\cos\beta$.
We have evaluated the cross section of Higgs bosons in $pp$ collisions
$\sigma(pp \to \phi +X)$, with two dominant subprocesses:
$gg \to \phi$ and $gg \to \phi b\bar{b}$.
For $m_A \agt$ 150 GeV, the couplings of the lighter scalar $h^0$
to gauge bosons and fermions become close to those of the SM Higgs boson,
therefore, gluon fusion is the major source of the $h^0$
even if $\tan\beta$ is large.


The QCD radiative corrections to $gg \to \phi$  
was found to be large \cite{QCD}, the same corrections 
to $gg \to \phi b\bar{b}$ are still to be evaluated.
To be conservative, we take a K-factor of 1.5 and 1.0 
for the contributions from $gg \to \phi$ 
and $gg \to \phi b\bar{b}$ respectively, 
to evaluate the cross section of $pp \to \phi +X$.
For the dominant Drell-Yan background \cite{Nikita,CMS,ATLAS2}, 
we have adopted the well known K-factor from reference \cite{Collider}.


If the $b\bar{b}$ mode dominates Higgs decays,
the branching fraction of $\phi \to \mu\bar{\mu}$ is about 
$m_\mu^2/3 m_b^2$, where 3 is the color factor of the quarks.
The QCD radiative corrections greatly reduce 
the decay width of $\phi \to b\bar{b}$ \cite{Manuel}.
For $\tan\beta \agt 10$, the $b\bar{b}$ decay mode dominates, 
and the branching fraction of 
$B(\phi \to \mu\bar{\mu})$ ($\phi = A^0, H^0$, or $h^0$) 
is about $2 \times 10^{-4}$.
For $m_A$ less than about 80 GeV, the $H^0$ decays dominantly into
$h^0 h^0$, $A^0 A^0$ and $Z A^0$.

\subsection*{Higgs Bosons of Minimal Supersymmetry}

In Figs. 3(a) and 3(b), 
we present the cross section of the MSSM Higgs bosons 
at the LHC, $pp \to \phi \to \mu\bar{\mu} +X$, as a function of $m_A$ 
for $\tan\beta = 15$ and $\tan\beta = 40$.
As $\tan\beta$ increases,
the cross section is enhanced because for $\tan\beta \agt 10$,
it is dominated by $gg \to \phi b\bar{b}$
and enhanced by the $\phi b\bar{b}$ Yukawa coupling.
Also shown is the same cross section for 
the SM Higgs boson $h^0_{SM}$ with $m_{h_{SM}} = m_A$.
For $m_{h_{SM}} > 140$ GeV, 
the SM $h^0_{SM}$ mainly decays into gauge bosons; 
therefore, the branching fraction $B(h^0_{SM} \to \mu\bar{\mu})$ 
drops sharply. 

\begin{figure}[ht] 
\centerline{\epsfig{file=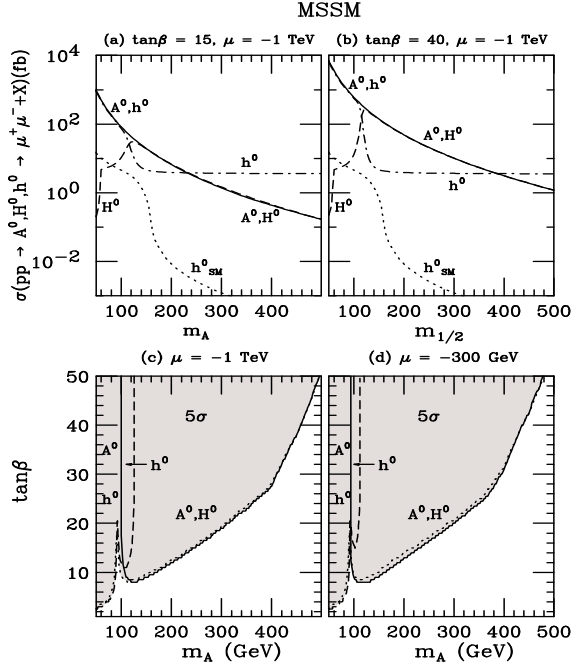,height=3.5in,width=3.0in}}
\vspace{10pt}
\caption{
The cross sections of $pp \to A^0,H^0,h^0 \to \mu\bar{\mu} +X$ in fb 
at $\sqrt{s} = 14$ TeV, versus $m_A$ 
for $m_{\tilde{g}} = m_{\tilde{q}} = -\mu = 1$ TeV, 
(a) $\tan\beta = 15$ and (b) $\tan\beta = 40$.
Also shown is the cross section for the SM Higgs boson 
with $m_{h_{SM}} = m_A$. 
The 5$\sigma$ contours at the LHC with $L =$ 300 fb$^{-1}$ 
are shown for 
(c) $m_{\tilde{g}} = m_{\tilde{q}} = -\mu = 1$ TeV, and 
(d) $m_{\tilde{g}} = m_{\tilde{q}} = -\mu = 300$ GeV. 
}
\label{figure3}
\end{figure}

To study the observability for the muon pair decay mode, 
the dominant background from the Drell-Yan (DY) process, 
$q\bar{q} \to Z,\gamma \to \mu\bar{\mu}$ is considered. 
We take $\Delta M_{\mu\bar{\mu}}$ to be the larger of
the ATLAS muon mass resolution (about $2\%$ of the Higgs bosons mass) 
\cite{ATLAS,ATLAS2} or the Higgs boson width.\footnote{
The CMS mass resolution will be better than $2\%$ of $m_\phi$ 
for $m_\phi \alt$ 500 GeV \cite{Nikita,CMS}.} 
The minimal cuts applied are (1) $p_T(\mu) > 20$ GeV and
(2) $|\eta(\mu)| < 2.5$ for both the signal and background.

For $m_A \agt$ 130 GeV, $m_A$ and $m_H$ are almost degenerate 
while for $m_A \alt$ 100 GeV, $m_A$ and $m_h0$ are very close 
to each other \cite{Nikita,CMS}.
Therefore, we sum up the cross sections of the $A^0$ and the $h^0$ 
for $m_A \le 100$ GeV and those of the $A^0$ and the $H^0$ 
for $m_A > 100$ GeV, 


We define the signal to be observable if the $99\%$
confidence level upper limit on the background is smaller than the
corresponding lower limit on the signal plus background \cite{HGG,Brown},
namely,
\begin{eqnarray}
L(\sigma_s+\sigma_b)-N\sqrt{L(\sigma_s+\sigma_b)} & > &
L\sigma_b+N\sqrt{L\sigma_b} \nonumber \\
\sigma_s & > & \frac{N^2}{L}[1+2\sqrt{L\sigma_b}/N]
\end{eqnarray}
where $L$ is the integrated luminosity, and $\sigma_b$ is the background cross
section within a bin of width $\pm\Delta M_{\mu\bar{\mu}}$ centered
at $M_\phi$; $N = 2.32$ corresponds to a $99\%$ confidence level
and $N = 2.5$  corresponds to a 5$\sigma$ signal.

The 5$\sigma$ discovery contours at $\sqrt{s} =$ 14 TeV 
and $L = 300 \;\; {\rm fb}^{-1}$ are shown in Figs. 3(c) and 3(d) for 
$m_{\tilde{q}} = m_{\tilde{g}} = -\mu = 1$ TeV
and $m_{\tilde{q}} = m_{\tilde{g}} = -\mu = 300$ GeV. 
The discovery region of $H^0 \to \mu\bar{\mu}$ 
is slightly enlarged for a smaller $\mu$,
but the observable region of $h^0 \to \mu\bar{\mu}$ is slightly reduced 
because the lighter top squarks make the $H^0$ and the $h^0$ lighter 
and enhance the $H^0 b\bar{b}$ coupling 
while reduce the $h^0 b\bar{b}$ coupling.

\subsection*{Higgs Bosons of Minimal Supergravity}

In the minimal supergravity model (MSUGRA) \cite{SUGRA}, 
it is assumed that SUSY is broken in a hidden sector 
with SUSY breaking communicated to the observable sector 
through gravitational interactions, 
leading naturally to a common scalar mass ($m_0$), 
a common gaugino mass ($m_{1/2}$), a common trilinear coupling ($A_0$) 
and a common bilinear coupling ($B_0$) at the GUT scale. 
Through minimization of the Higgs potential, the $B$ parameter 
and magnitude of the superpotential Higgs mixing parameter $\mu$ 
are related to $\tan\beta$ and $M_Z$. 

The SUSY particle masses and couplings at the weak scale 
can be predicted by the evolution of RGEs \cite{RGE}
from the unification scale \cite{BBO,SUGRA2}. 
Since  $A_0$ mainly affects the masses of third generation sfermions, 
it is taken to be zero in most of our analysis. 
We calculate masses and couplings in the Higgs sector 
with one loop corrections from the top and the bottom Yukawa interactions 
in the RGE-improved one-loop effective potential \cite{Higgs} 
at the scale $Q = \sqrt{m_{\tilde{t}_L}m_{\tilde{t}_R}}$ \cite{Baer,Madison}. 
At this scale, the RGE improved one-loop corrections approximately 
reproduce the dominant two loop corrections \cite{Two-Loop}  
to the mass of the lighter CP-even scalar ($m_h$).

The mass matrix of the charginos in the weak eigenstates 
($\tilde{W}^\pm$, $\tilde{H}^\pm$) has the following form \cite{BBO}
\begin{equation}
M_C=\left( \begin{array}{c@{\quad}c}
M_2 & \sqrt{2}M_W\sin \beta \\
\sqrt{2}M_W\cos \beta & -\mu
\end{array} \right)\;.
\label{eq:xino}
\end{equation}
The form of Eq.~(\ref{eq:xino}) establishes our sign convention for $\mu$. 
Recent measurements of the $b \to s\gamma$ decay rate 
by the CLEO \cite{CLEO} and the LEP collaborations \cite{LEP} 
excludes most of the MSUGRA parameter space 
for $\mu > 0$ with a large $\tan\beta$ \cite{bsg}. 
Although we choose $\mu < 0$ in our analysis, 
our results and conclusions are almost independent of the sign of $\mu$.

Figure 4 shows masses, in the case of $\mu < 0$, 
for neutral Higgs bosons: the lighter CP-even ($h^0$), 
the heavier CP-even ($H^0$) and the  CP-odd ($A^0$). 
Also shown are the regions that do not satisfy 
the following theoretical requirements: 
electroweak symmetry breaking (EWSB), tachyon free, 
and the lightest neutralino ($\chi^0_1$) 
as the lightest supersymmetric particle (LSP).
The region excluded by the $m_{\chi^+_1} > 85$ GeV limit 
from the chargino search \cite{ALEPH} at LEP 2 is indicated.
There are a couple of interesting aspects to note: 
(i) an increase in $\tan\beta$ leads to a larger $m_h$ 
but a reduction in $m_A$ and $m_H$; 
(ii) increasing $m_0$ raises $m_A$, $m_H$ and masses of the other scalars 
significantly. 

\begin{figure}[ht] 
\centerline{\epsfig{file=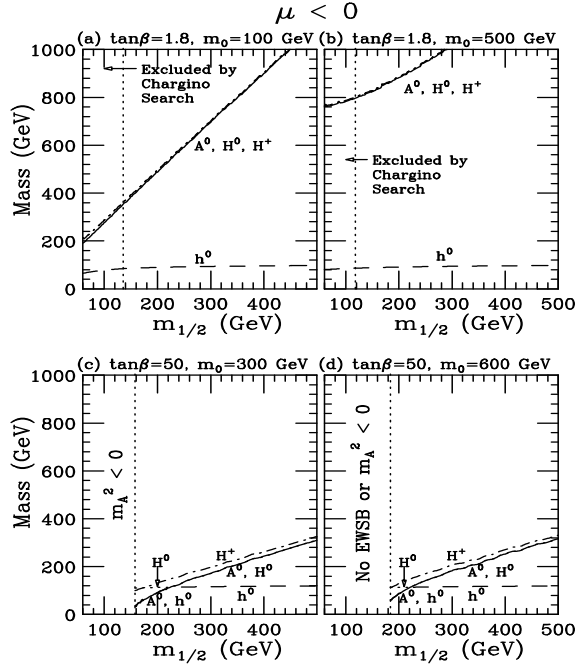,height=3.5in,width=3.0in}}
\vspace{10pt}
\caption{
Masses of $H^0$, $h^0$, and $A^0$ at the mass scale
$Q = \sqrt{  m_{\tilde{t}_L} m_{\tilde{t}_R} }$,
versus $m_{1/2}$.
}
\label{figure4}
\end{figure}

The LHC discovery contours in the minimal supergravity model 
are presented in Figure 5 for 
(a) the $m_{1/2}$ versus $\tan\beta$ plane with $m_0 = 150$ GeV,
(b) the $m_{1/2}$ versus $\tan\beta$ plane with $m_0 = 500$ GeV,
(c) the $m_{1/2}$ versus $m_0$ plane with $\tan\beta = 15$, and
(d) the $m_{1/2}$ versus $m_0$ plane with $\tan\beta = 40$.
The discovery region is the part of the parameter space 
between the curve of square symbol and the dash line. 
The QCD radiative corrections to background from the Drell-Yan process
are included.

\begin{figure}[t] 
\centerline{\epsfig{file=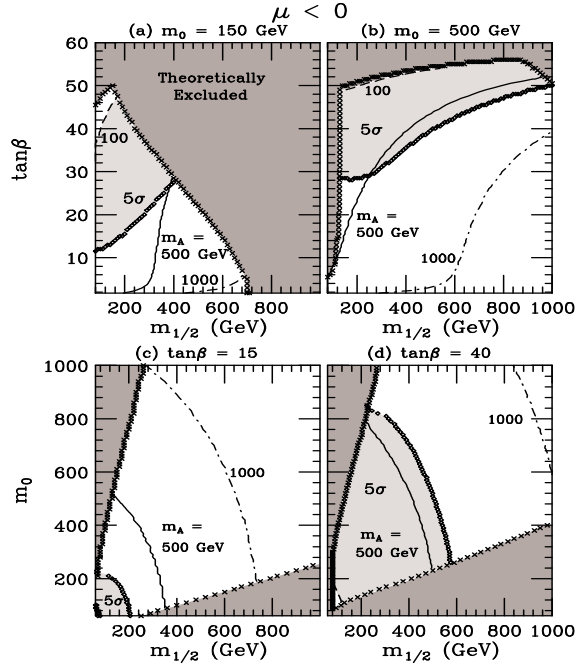,height=3.5in,width=3.0in}}
\vspace{10pt}
\caption{
The $5\sigma$ contours for detecting Higgs bosons of MSUGRA at the LHC 
with $L =$ 300 fb$^{-1}$.
Also shown are 
(i) the mass contours for $m_A = 100$ GeV, 500 GeV and 1000 GeV, 
(ii) the parts of the parameter space excluded 
by theoretical requirements (dark shading), and  
(iii) the region excluded by the $m_{\chi^+_1} > 85$ GeV limit 
from the chargino search at LEP 2.
}
\label{figure5}
\end{figure}

\section*{Conclusions}

The discovery  channel of $A^0 \to \gamma \gamma$ 
might provide a good opportunity to precisely reconstruct 
the CP-odd Higgs boson mass ($m_A$) for 170 GeV $< m_A < 2m_t$ 
if the decays of the $A^0$ into SUSY particles
are forbidden and $\tan\beta$ is close to one.
The impact of SUSY decays on this discovery channel 
might be significant \cite{Z2Z2} 
and it is under investigation with realistic simulations.

The muon pair decay mode can be a very promising channel
to discover the neutral Higgs bosons of minimal supersymmetry 
and minimal supergravity, and this mode will provide a good channel 
to precisely reconstruct Higgs boson masses.
The $A^0$ and $H^0$ might be observable in a large region of parameter space
with $\tan\beta \agt 10$.
The $h^0$ might be observable in a region with $m_A < 120$ GeV
and $\tan\beta \agt 5$.
For $m_A \agt 200$ GeV and $\tan\beta > 25$, $L =$ 10 fb$^{-1}$
would be enough to obtain Higgs boson signals
with a statistical significance larger than 7 \cite{Nikita}. 

In the MSUGRA, the observable regions of the parameter space are found to be
\begin{eqnarray}
m_0 = 150 \;\; {\rm GeV}: & & \;\;\;\, m_{1/2} \alt 400 \;\;{\rm GeV}
\;\; {\rm and} \;\; \tan\beta \agt 12 \nonumber \\
m_0 = 500 \;\; {\rm GeV}: & & \;\;\;\, m_{1/2} \alt 1 \;\;{\rm TeV}
\;\; {\rm and} \;\; \tan\beta \agt 28
\end{eqnarray}
For two specific choices of large $\tan\beta$, the observable regions are
\begin{eqnarray}
\tan\beta = 15: & & \;\;\;\, m_{1/2} \alt 200 \;\;{\rm GeV}
\;\; {\rm and} \;\; m_0 \alt 200 \;\; {\rm GeV} \nonumber \\
\tan\beta = 40: & & \;\;\;\, m_{1/2} \alt 600 \;\;{\rm GeV}
\;\; {\rm and} \;\; m_0 \alt 800 \;\; {\rm GeV}.
\end{eqnarray}

\section*{Acknowledgments}

I am grateful to Salavat Abdullin, Vernon Barger and Nikita Stepanov 
for enjoyable and inspiring collaborations. 
This research was supported in part by the U.S. Department of Energy
under Grant No. DE-FG02-95ER40896,
and in part by the University of Wisconsin Research Committee
with funds granted by the Wisconsin Alumni Research Foundation.

\end{document}